\documentclass[aps, prl,superscriptaddress,
amsmath,amssymb,preprintnumbers, twocolumn, showkeys, showpacs]{revtex4-1}

\usepackage{graphicx}
\usepackage{dcolumn}
\usepackage{bm}

\begin{document}


\title{The Gaussian-Smoothed Wigner Function and Its Application to Precision Analysis}

\author{Hai-Woong Lee}
\email{hwlee@unist.ac.kr}
\affiliation{Division of General Studies, Ulsan National Institute of Science and Technology, Ulsan 689-798, Republic of Korea}
\affiliation{Department of Physics, Korea Advanced Institute of Science and Technology, Daejeon 305-701, Republic of Korea}

\date{\today}

\pacs{03.65.Ta, 42.50.Dv}
\keywords{Wigner function, Gaussian convolution, eight-port homodyne detection}

\begin{abstract}
We study a class of phase-space distribution functions that is generated from a Gaussian convolution of the Wigner distribution function.
This class of functions represents the joint count probability in simultaneous measurements of position and momentum. We show that, using
these functions, one can determine the expectation value of a certain class of operators accurately, even if measurement data performed
only with imperfect detectors are available. As an illustration, we consider the eight-port homodyne detection experiment that performs
simultaneous measurements of two quadrature amplitudes of a radiation field.
\end{abstract}

\maketitle
The phase-space formulation of quantum mechanics, which originates from the classic work of Wigner\cite{Wigner}, has enjoyed a wide popularity
in all areas of physics\cite{Lee, Schleich}. As there is no unique way of assigning a quantum-mechanical operator to a given classical function
of conjugate variables, there can exist many different quantum phase-space distribution functions, the best known of which are the Wigner
function\cite{Wigner}, Husimi function\cite{Husimi}, and P and Q functions\cite{Glauber, Sudarshan}. All these distribution functions are
equivalent to one another, in the sense that any of them can be used to evaluate the expectation value of any arbitrary operator. Only the rule
of ordering of noncommuting operators is different for a different function.

In this work, we study a class of distribution functions that results from a Gaussian convolution of the Wigner function in phase space. The
starting point of our study is the equation
\begin{equation}
H(q,p)=\frac{1}{\pi \hbar} \int dq^{\prime} \int dp^{\prime} \ e^{- \frac{(q^{\prime}-q)^2 }{2\sigma_{q}^{2}}} e^{-\frac{(p^{\prime}-p)^2 }
{2(\hbar /2\sigma_q )^2 }} \ W(q^{\prime} ,p^{\prime} ) ,
\end{equation}
which defines the Husimi function $H$ in terms of a Gaussian convolution of the Wigner function $W$ in phase space \cite{Lee,Lee2}. Note that
the width $\sigma_q $ of the Gaussian function in $q$ space and the width $\sigma_p =\hbar /2 \sigma_q $ in $p$ space satisfy the Heisenberg
minimum uncertainty relation
\begin{equation}
\sigma_q \sigma_p =\frac{\hbar}{2} .
\end{equation}
It is straightforward to generalize Eq.(1) and define the Gaussian-smoothed Wigner function $G$ as
\begin{equation}
G(q,p)=\frac{1}{2\pi \sigma_q \sigma_p } \int dq^{\prime} \int dp^{\prime}\ e^{- \frac{(q^{\prime}-q)^2 }{2\sigma_{q}^{2}}} e^{-\frac{(p^{\prime}-p)^2 }
{2 \sigma_{p}^{2} }} \ W(q^{\prime} ,p^{\prime} ) .
\end{equation}

The Husimi function is everywhere real and nonnegative\cite{Cartwright} and is entitled to probability interpretation. It has
indeed been shown that the Husimi function represents a proper probability distribution associated with ideal simultaneous measurements
of position and momentum\cite{AK, BCM, Stenholm}, where the ideal measurement refers to a measurement performed with a perfect measurement
device satisfying the Heisenberg minimum uncertainty relation. We note that the widths $\sigma_q $ and $\sigma_p $ of the smoothing Gaussian
function in Eq. (1) are identified with the measurement uncertainties in $q$ and $p$, respectively. In other words, the act of simultaneous
measurement is modeled by phase-space Gaussian smoothing, with the widths of the smoothing Gaussian function identified as measurement
uncertainties\cite{Wodkiewicz}. The physical significance of the function $G$ defined by Eq. (3) should now be clear. It represents a probability
distribution resulting from simultaneous measurements of position and momentum, where the measurements are performed with a device
characterized by measurement uncertainties $\sigma_q $ and $\sigma_p $. One may consider a large number of identically prepared systems on
each of which a simultaneous measurement of position and momentum is performed. Each time, the measurement is performed with an identical
measurement device of measurement uncertainties $\sigma_q $ and $\sigma_p $. The probability distribution in $q$ and $p$ resulting from such
measurements is the function $G$ of Eq. (3), where the widths $\sigma_q $ and $\sigma_p $ of the smoothing Gaussian function are given by
the measurement uncertainties in $q$ and $p$, respectively, of the measurement device used. According to the Heisenberg uncertainty principle,
the measurement uncertainties must satisfy
\begin{equation}
\sigma_q \sigma_p \geq \frac{\hbar}{2} .
\end{equation}
The function $G$, with $\sigma_q \sigma_p \geq \frac{\hbar}{2} $, is thus physically measurable through simultaneous measurements of
position and momentum.

It should be emphasized that, at least in principle, the function $G$ is as good a quantum phase-space distribution function as the Husimi
function or the Wigner function. The expectation value of any arbitrary operator can be calculated using the function $G$ as well as
the Husimi function or the Wigner function. Only the rule of ordering of noncommuting operators is different. In order to find the rule of
ordering associated with the function $G$, we begin with the equation\cite{Lee}
\begin{equation}
Tr \{ \hat{\rho} \ e^{i \xi \hat{q} +i \eta \hat{p}} f(\xi, \eta) \}=\int dq \int dp \ e^{i \xi q+i \eta p} \ G(q,p) .
\end{equation}
It can be shown that the function $f(\xi,\eta)$ that determines the rule of ordering for the function $G$ is given by
\begin{equation}
f(\xi, \eta)=e^{-\sigma_{q}^{2} \xi^{2} /2-\sigma_{p}^{2} \eta^{2} /2 } .
\end{equation}
At this point we find it convenient to introduce two parameters $\kappa$ and $s$ defined as
\begin{equation}
\kappa =\frac{\sigma_p}{\sigma_q}
\end{equation}
and
\begin{equation}
s=-\frac{\sigma_q \sigma_p}{\hbar /2} =-\frac{\kappa \sigma_{q}^{2}}{\hbar /2} .
\end{equation}
The parameter $\kappa$ has a dimension of $m \omega$ [mass/time]. The parameter $s$ is real and negative, and its absolute
value measures the product of the widths $\sigma_q$ and $\sigma_p$ associated with the function $G$ being considered with respect to
that of the minimum uncertainty Gaussian wave packet. Once $\sigma_q$ and $\sigma_p$ are given, $\kappa$ and $s$ are determined, and
vice versa. Eq. (6) can be rewritten, in terms of $\kappa$ and $s$, as
\begin{equation}
f(\xi,\eta)=e^{s \frac{\hbar \xi^{2} }{4 \kappa}+s \frac{\hbar \kappa \eta^{2}}{4}} .
\end{equation}
We further introduce dimensionless parameters $v$ and $\beta$ and an operator $\hat{b}$ as
\begin{equation}
v=i \xi \sqrt{\frac{\hbar}{2 \kappa}} -\eta \sqrt{\frac{\hbar \kappa}{2}} \ ,
\end{equation}
\begin{equation}
\beta =\sqrt{\frac{\kappa}{2 \hbar}}q+\frac{i}{\sqrt{2\hbar \kappa}} p \ ,
\end{equation}
\begin{equation}
\hat{b} =\sqrt{\frac{\kappa}{2 \hbar}}\hat{q}+\frac{i}{\sqrt{2\hbar \kappa}} \hat{p} \ .
\end{equation}
Eq. (5) can then be rewritten as
\begin{eqnarray}
Tr \{\hat{\rho} \ e^{v {\hat{b}}^{\dagger} -v^{\ast} \hat{b} } e^{s \frac{|v|^2 }{2} } \}
=Tr \{ \hat{\rho} \ e^{-v^{\ast} \hat{b}} e^{v {\hat{b}}^{\dagger}} e^{(s+1) \frac{|v|^2 }{2}} \} \nonumber \\
=\int d^{2} \beta \ e^{v \beta^{\ast} -v^{\ast} \beta} \ G(\beta, \beta^{\ast}) .
\end{eqnarray}
The rule of ordering for the function $G$ can now be determined from Eq. (13) using the same method that Cahill and Glauber\cite{CG}
adopted for their s-parameterized distribution function. The final result is
\begin{equation}
\{ \hat{b} ^{\dagger n} \hat{b} ^{m} \} =\sum_{k=0}^{(n,m)} k! \left( \begin{array}{c} n \\ k \end{array} \right)
\left( \begin{array}{c} m \\ k \end{array} \right) (-\frac{s}{2} -\frac{1}{2} )^{k} \ \hat{b}^{(m-k)} \hat{b}^{\dagger (n-k)} ,
\end{equation}
where $\{ \hat{b}^{\dagger n} \hat{b} ^{m} \}$ represents the rule of ordering for the function $G$, the symbol $(n,m)$ denotes
the smaller of the two integers $n$ and $m$, and $\left( \begin{array}{c} n \\ k \end{array} \right)$ is a binomial coefficient.
Eq. (14) yields, for example, $\{\hat{b}^{\dagger} \}=\hat{b}^{\dagger} $, $\{ \hat{b} \} = \hat{b}$, $\{\hat{b}^{\dagger} \hat{b} \}
=\hat{b} \hat{b}^{\dagger} -\frac{1}{2} (s+1) $, $\{ \hat{b}^{\dagger} \hat{b}^{2} \}=\hat{b}^{2} \hat{b}^{\dagger} -(s+1) \hat{b}$,
$\{ \hat{b}^{\dagger 2} \hat{b} \}=\hat{b} \hat{b}^{\dagger 2} -(s+1) \hat{b}^{\dagger}$, and $\{ \hat{b}^{\dagger 2} \hat{b}^{2} \}
=\hat{b}^{2} \hat{b}^{\dagger 2}-2(s+1)\hat{b} \hat{b}^{\dagger} +\frac{1}{2} (s+1)^{2} $.

As an illustration, let us find the expectation value of $\hat{q} \hat{p}^2$ with the function $G(q,p)$. We first need to use
Eqs. (7) and (11) and perform change of variables from $(q, p)$ to $(\beta, \beta^{\ast}) $ to obtain $G(\beta, \beta^{\ast}) $.
Expressing $\hat{q} \hat{p}^{2}$ in terms of $\{ \hat{b}^{\dagger m} \hat{b}^{n} \}$, we obtain
\begin{eqnarray}
\hat{q} \hat{p}^{2} =-\sqrt{\frac{\hbar^{3} \kappa}{8}}[\{\hat{b}^{3} \}+\{\hat{b}^{\dagger 3} \}-\{ \hat{b}^{\dagger} \hat{b}^{2} \}
-\{ \hat{b}^{\dagger 2} \hat{b} \} \nonumber \\
-(s+2)\{\hat{b} \}-(s-2)\{ \hat{b}^{\dagger} \} ] .
\end{eqnarray}
Eq. (15) leads immediately to
\begin{eqnarray}
\hat{q} \hat{p}^{2} =-\sqrt{\frac{\hbar^{3} \kappa}{8}} \int d^{2} \beta \ G(\beta, \beta^{\ast})   [\beta^3 +\beta^{\ast 3} \nonumber \\
-\beta^2 \beta^{\ast}
-\beta \beta^{\ast 2}
-(s+2)\beta -(s-2)\beta^{\ast}].
\end{eqnarray}

We now discuss a possible application of the function $G$ to precision analysis. When performing measurements, one faces a realistic
problem of having to deal with imperfect detectors of efficiencies lower than 1. For precision analysis, one needs to find a reliable
way of correcting unavoidable errors arising from imperfect measurements. An attractive feature of the function $G$ introduced here
is that it provides a definite recipe, in the form of the rule of ordering, that allows one to obtain precise quantitative information about the
system being considered from measurement data collected by imperfect detectors. Let us suppose that simultaneous measurements of two
conjugate variables are performed with realistic detectors of efficiencies less than 1, from which the joint count probability
distribution is obtained. One can identify exactly the function $G$ representing this distribution, as long as the efficiencies of
the detectors are known. Since the rule of ordering of noncommuting operators for the function $G$ is exactly known, the expectation
value of any operator can be evaluated using the function $G$. Hence, at least in principle, a high degree of accuracy comparable to
that with near-perfect detectors is within reach, even if measurements are performed with imperfect detectors. We illustrate this below
by considering the eight-port homodyne detection experiment\cite{Schleich, NFM} that performs simultaneous measurements of two
quadrature amplitudes of a radiation field.

When applied to a radiation field, which mathematically is equivalent to a harmonic oscillator of mass $m=1$, Eq. (3) translates into
\begin{equation}
G(\alpha, \alpha^{\ast})=
\frac{1}{2\pi \sigma_1 \sigma_2 } \int d^{2} \alpha^{\prime} \ e^{-\frac{(\alpha_{1}^{\prime}-\alpha_{1} )^2 }{2\sigma_{1}^{2}}}
e^{-\frac{(\alpha_{2}^{\prime} -\alpha_{2} )^2 }{2 \sigma_{2}^{2}}} W(\alpha^{\prime}, \alpha^{\ast \prime} ),
\end{equation}
where $\alpha_1$ and $\alpha_2$, real and imaginary parts of $\alpha$, respectively, refer to two quadrature amplitudes of the radiation
field. When $\sigma_1 \sigma_2 =\frac{1}{4}$, the function $G$ becomes the Husimi function. In particular, when $\sigma_1 =\sigma_2 =
\frac{1}{2}$, the function $G$ is reduced to the Q function. Furthermore, when $\sigma_1 =\sigma_2 \equiv \sigma$, the function $G$
becomes the s-parameterized function of Cahill and Glauber\cite{CG} with $s=-4\sigma^{2}$.

Simultaneous measurements of two quadrature amplitudes of a radiation field, contingent upon the Heisenberg uncertainty principle
$\sigma_1 \sigma_2 \geq \frac{1}{4}$, can be performed using the eight-port homodyne detection scheme proposed earlier\cite{NFM}.
It has been shown\cite{FVS, LP} that the joint count probability in the two detectors used for the scheme is given, in the limit
of a strong local oscillator, by the Q function of the signal field, provided that the two detectors are perfect. In reality,
however, detectors have nonunit efficiency $\eta <1$. Assuming that the two detectors have the identical nonunit efficiency
$\eta$, Leonhardt and Paul\cite{LP2} have shown that the eight-port homodyne scheme measures the s-parameterized function of Cahill
and Glauber\cite{CG}, where $s=-\frac{2-\eta}{\eta}$. This result can be generalized in a straightforward way to the case where
the two detectors have different efficiencies $\eta_1$ and $\eta_2$. In this case, the eight-port homodyne scheme measures the
function $G$ with $\sigma_1$ and $\sigma_2$ given by $4\sigma_{1}^{2} =-s_1 =\frac{2-\eta_1 }{\eta_1 }$ and $4\sigma_{2}^{2} =
-s_2 =\frac{2-\eta_2 }{\eta_2 }$, respectively. For this case, the parameters $\kappa$ and $s$ are given by (with mass $m=1$)
\begin{equation}
\kappa=\omega \frac{\sigma_2}{\sigma_1} =\omega \sqrt{\frac{(2-\eta_2 )\eta_1 }{(2-\eta_1 )\eta_2 }},
\end{equation}
\begin{equation}
s=-4\sigma_1 \sigma_2 =-\sqrt{\frac{(2-\eta_1 )(2-\eta_2 )}{\eta_1 \eta_2 }},
\end{equation}
where $\omega$ is the angular frequency of the field. The corresponding rule of ordering is given by Eq. (14), where the operator
$\hat{b}$, the annihilation operator of a ''squeezed'' photon, is defined in Eq. (12) with $\kappa$ given by Eq. (18).

For example, let us suppose that we wish to find the expectation value of the photon number, $\langle \hat{a}^{\dagger} \hat{a}
\rangle$, in the signal field. For this purpose, we first need to express $\langle \hat{a}^{\dagger} \hat{a} \rangle$ in
terms of $\{ \hat{b}^{\dagger n} \hat{b} ^{m} \}$ of Eq. (14). A straightforward algebra yields
\begin{equation}
\hat{a}^{\dagger} \hat{a}=-(\{\hat{b}^{\dagger 2} \}+\{\hat{b}^{2} \})\frac{sinh 2r}{2} +\{\hat{b}^{\dagger}\hat{b} \} cosh 2r
+\frac{s}{2} cosh 2r -\frac{1}{2},
\end{equation}
where the ''squeeze'' parameter $r$ is defined as
\begin{equation}
e^r =\sqrt{\frac{\kappa}{\omega}}.
\end{equation}
We thus have
\begin{eqnarray}
\langle \hat{a}^{\dagger} \hat{a} \rangle=\int d^2 \beta \ G(\beta, \beta^{\ast} )[-(\beta^{\ast 2} +\beta^{2} )
\frac{sinh 2r}{2} \nonumber \\
+|\beta|^{2} cosh 2r ]
+\frac{s}{2} cosh 2r -\frac{1}{2}.
\end{eqnarray}
Here, the parameter $\beta$ is related to the quadrature amplitude $\alpha$ by
\begin{equation}
\beta=\alpha cosh r+\alpha^{\ast} sinh r
\end{equation}
The joint count probability of the two (imperfect) detectors in the eight-port homodyne scheme leads directly to the identification
of the function $G(\alpha, \alpha^{\ast} )$. One can then obtain $G(\beta, \beta^{\ast} )$ through change of variables from
$(\alpha, \alpha^{\ast})$ to $(\beta, \beta^{\ast})$. The expectation value $\langle \hat{a}^{\dagger} \hat{a} \rangle$ can then be
calculated using Eq. (22).

We emphasize that Eq. (22) is valid for any arbitrary values $\eta_1 $ and $\eta_2 $ of the efficiencies of the detectors used.
Hence, an accurate determination of $\langle \hat{a}^{\dagger} \hat{a} \rangle $ can be achieved even from measurement data performed
with imperfect detectors. One only needs to determine the function $G(\alpha, \alpha^{\ast}) $ accurately from the joint count probability
of the (imperfect) detectors used. The function to be determined here is the function $G$ associated with the very detectors used, not
the function $G(=Q)$ which would be obtained with the ideal detectors.
To elaborate further on this point, let us consider the simple case when the two
detectors have the same efficiency $\eta_1 =\eta_2 \equiv \eta$. In this case, we have $\kappa=\omega$, $\beta=\alpha$, and
$r=0$, and Eq. (22) is simplified to
\begin{equation}
\langle \hat{a}^{\dagger} \hat{a} \rangle =\int d^2 \alpha \ G(\alpha, \alpha^{\ast})|\alpha|^{2} +\frac{1}{2} (s-1).
\end{equation}
If the measurements were performed with the perfect detectors, the joint count probability would yield the Q function\cite
{FVS, LP}, and the expectation value $\langle \hat{a}^{\dagger} \hat{a} \rangle $ would be calculated by
\begin{equation}
\langle \hat{a}^{\dagger} \hat{a} \rangle =\langle \hat{a} \hat{a}^{\dagger} \rangle -1=
\int d^2 \alpha \ Q(\alpha, \alpha^{\ast})|\alpha|^{2} -1.
\end{equation}
The difference between Eq. (24) and (25), namely $\frac{1}{2} (s+1)$, represents the ''correction'' factor that needs to be
added to compensate for the use of imperfect detectors. This presents no problem, because the correction factor can be
determined exactly [see Eq. (19)] once the efficiency $\eta$ is known. One can thus say that the expectation value
 $\langle \hat{a}^{\dagger} \hat{a} \rangle $ can be determined from the eight-port homodyne detection experiment to a
 near-perfect degree of accuracy, regardless of the efficiencies of the detectors used.

 Difficulty arises, however, when one wants to evaluate the expectation values $\langle \hat{a}^{\dagger n} \hat{a}^{m} \rangle $
 for large integers $n$ and $m$. In general, the determination of $\langle \hat{a}^{\dagger n} \hat{a}^{m} \rangle $
 requires an accurate evaluation of the integrals $\int d^2 \beta \ G(\beta, \beta^{\ast}) \beta^{\ast n} \beta^{m} $,
 $\int d^2 \beta \ G(\beta, \beta^{\ast}) \beta^{\ast n-1} \beta^{m-1} $, etc. When $n$ and/or $m$ are large, the value of
 these integrals may vary widely with respect to small changes in the function $G$, and thus it is important to determine
 accurately the function $G$ from the experiment. When low-efficiency detectors are used, the function $G$ resulting from
 the joint count probability is a strongly smoothed function and thus exhibits a relatively flat distribution. In such
 a case, an accurate determination of the function $G$ requires an accurate dteremination of a large number of significant
 figures of its values, which puts a heavy (and perhaps impossible if $n$ and $m$ are quite large and if the detector
 efficiencies deviate significantly from unity) burden on the experiment. Another difficulty arises from the fact that an
 accurate evaluation of $\langle \hat{a}^{\dagger n} \hat{a}^{m} \rangle $ requires an accurate evaluation of lower-order
 expectation values, and thus a small error in $\langle \hat{a}^{\dagger} \hat{a} \rangle $ and other low-order expectation
 values are magnified in the evaluation of
 high-order expectation values $\langle \hat{a}^{\dagger n} \hat{a}^{m} \rangle $. One thus concludes that, the greater
 the integers $n$ and $m$ are, the closer to unity the efficiencies $\eta_1 $ and $\eta_2 $ are required to be for an accurate
 evaluation of $\langle \hat{a}^{\dagger n} \hat{a}^{m} \rangle $, i.e., the requirement on detector efficiencies gets
 increasingly severe for larger integer values of $n$ and $m$. In other words, the lower the efficiencies of the detectors
 used are, the more strongly limited the number of expectation values $\langle \hat{a}^{\dagger n} \hat{a}^{m} \rangle $
 that can be determined reliably. With imperfect detectors, it is practically impossible to accurately evaluate the
 expectation values $\langle \hat{a}^{\dagger n} \hat{a}^{m} \rangle $ for all integers $n$ and $m$. Hence, an accurate
 state reconstruction, for example, which requires $\langle \hat{a}^{\dagger n} \hat{a}^{m} \rangle $ for all integers
 $n$ and $m$, may be difficult, unless measurements are performed with near-perfect detectors. Nevertheless, if one is
 primarily interested in the expectation values $\langle \hat{a}^{\dagger n} \hat{a}^{m} \rangle $ for small integers
 $n$ and $m$, then the phase-space approach with the function $G$ may provide a way to determine them with a high degree
 of accuracy, even if one is equipped only with moderately imperfect detectors.

 The question arises how low the efficiencies can be if one can still hope to get reliable quantitative information about
 the system being measured. As a rough estimate, it may be expected that the function $G$ with $\sigma_{1}^{2} \approx
 \sigma_{2}^{2} \approx \frac{1}{2} $ is as good a phase-space distribution function as the Q function, in the same sense
 that the Q function with $\sigma_{1}^{2} \approx \sigma_{2}^{2} \approx \frac{1}{4} $ is as good a function as the
 Wigner function. Taking $\sigma_{1}^{2} = \sigma_{2}^{2} = \frac{1}{2} $, one obtains $\eta_1 =\eta_2 \approx0.67$.
 If the two detectors used in the eight-port homodyne detection experiment have efficiencies higher than $\sim 0.67$, then
 the function $G$ constructed from the experiment may be expected to provide reasonably accurate information about the
 signal field.

 The difficulty mentioned above associated with a strongly smoothed function $G$ derived from low-efficiency detectors
 has its mathematical root in the fact
 that the Gaussian convolution operation of Eq. (3) [or Eq. (17)] is, as has already been noted\cite{AW}, the two-dimensional
 Weierstrass transform, which is an invertible point-to-point integral transform. As such, there is, in principle, no information
 loss when the convolution operation is performed, even if fine structures are inevitably smoothed. This is consistent with
 the fact that, regardless of the strength of smoothing, a definite rule of ordering exists in the form of Eq. (14),
 which enables, in principle, an accurate evaluation of the expectation values $\langle \hat{a}^{\dagger n} \hat{a}^{m} \rangle $
 for all integers $n$ and $m$. There, however, exists practical difficulty with the inverse Weierstrass transform, because
 a small error is magnified exponentially in the inverse transform. Hence, the requirement on the accuracy of the function $G$
 is increasingly severe, as the strength of smoothing is increased. Despite this practical difficulty, it is still encouraging
 that the expectation values $\langle \hat{a}^{\dagger n} \hat{a}^{m} \rangle $ for small integers can be accurately evaluated,
 even if one has only imperfect detectors and is therefore provided with a strongly smoothed function $G$.

 In summary, we have found the rule of ordering of conjugate variables for the Gaussian-smoothed Wigner function $G$, which
 allows an accurate evaluation of the expectation values $\langle \hat{q}^{n} \hat{p}^{m} \rangle $ (or $\langle \hat{a}^{\dagger n}
 \hat{a}^{m} \rangle $). On the basis of the fact that the function $G$ represents the joint count probability in simultaneous
 measurements of two conjugate variables $q$ and $p$ (or $\alpha_1 $ and $\alpha_2$), we have shown that the data obtained from
 simultaneous measurements performed with realistic, nonideal detectors can be analyzed in such a way that a fairly accurate
 evaluation of the expectation values $\langle \hat{q}^{n} \hat{p}^{m} \rangle $ (or $\langle \hat{a}^{\dagger n}
 \hat{a}^{m} \rangle $) for low intergers $n$ and $m$ can be achieved.



\begin{thebibliography}{999}


\bibitem{Wigner}
 E. Wigner, Phys. Rev. {\bf 40}, 749 (1930).
 \bibitem{Lee}
 H. W. Lee, Phys. Rep. {\bf 259}, 147 (1995).
 \bibitem{Schleich}
 W. P. Schleich, ''{\it Quantum Optics in Phase Space}'' (Wiley-VCH, Berlin 2001).
 \bibitem{Husimi}
 K. Husimi, Proc. Phys. Math. Soc. Japan {\bf 22}, 264 (1940).
 \bibitem{Glauber}
 R. J. Glauber, Phys. Rev. {\bf 131}, 2766 (1963).
 \bibitem{Sudarshan}
 E. C. G. Sudarshan, Phys. Rev. Lett. {\bf 10}, 277 (1963).
\bibitem{Lee2}
 H. W. Lee, Phys. Rev. A {\bf 50}, 2746 (1999).
 \bibitem{Cartwright}
 N. D. Cartwright, Physica A {\bf 83}, 210 (1976).
 \bibitem{AK}
 E. Arthurs and J. L. Kelly, Jr., Bell Syst. Tech. J. {\bf 44}, 725 (1965).
 \bibitem{BCM}
 S. L. Braunstein, C. M. Caves and G. J. Milburn, Phys. Rev. A {\bf 43}, 1153 (1991).
 \bibitem{Stenholm}
 S. Stenholm, Ann. Phys. {\bf 218}, 233 (1992).
 \bibitem{Wodkiewicz}
 K. Wodkiewicz, Phys. Rev. Lett. {\bf 52}, 1064 (1984).
 \bibitem{CG}
 K. E. Cahill and R. J. Glauber, Phys. Rev. {\bf 177}, 1857 (1969).
 \bibitem{NFM}
 J. W. Noh, A. Fougeres and L. Mandel, Phys. Rev. Lett. {\bf 67}, 1426 (1991).
 \bibitem{FVS}
 M. Freyberger, K. Vogel and W. P. Schleich, Phys. Lett. A {\bf 176}, 41 (1993).
 \bibitem{LP}
 U. Leonhardt and H. Paul, Phys. Rev. A {\bf 47}, R2460 (1993).
 \bibitem{LP2}
 U. Leonhardt and H. Paul, Phys. Rev. A {\bf 48}, 4598 (1993).
 \bibitem{AW}
 G. S. Agarwal and E. Wolf, Phys. Rev. D {\bf 2}, 2161 (1970).


\end{thebibliography}
\end{document}